\documentclass[aps,superscriptaddress]{revtex4}
\usepackage{amsmath}
\usepackage{graphicx}
\usepackage{amsfonts}
\usepackage{amssymb}

\begin{document}

\title{Chaotic Dynamics of a Free Particle Interacting Linearly with a
Harmonic Oscillator}
\author{Stephan De Bi\`{e}vre}
\affiliation{Laboratoire P.Painlev\'{e} and UFR de Math\'{e}matiques, Universit\'{e} des
Sciences et Technologies de Lille, 59655 Villeneuve d'Ascq, FRANCE}
\author{Paul E. Parris}
\affiliation{Laboratoire P.Painlev\'{e} and UFR de Math\'{e}matiques, Universit\'{e} des
Sciences et Technologies de Lille, 59655 Villeneuve d'Ascq, FRANCE}
\affiliation{Department of Physics, University of Missouri-Rolla, Rolla, MO 65409}
\affiliation{Consortium of the Americas for Interdisciplinary Science and Department of
Physics and Astronomy, University of New Mexico, Albuquerque, NM 87131}
\author{Alex Silvius}
\affiliation{Department of Physics, University of Missouri-Rolla, Rolla, MO 65409}

\begin{abstract}
We study the closed Hamiltonian dynamics of a free particle moving on a
ring, over one section of which it interacts linearly with a single harmonic
oscillator. On the basis of numerical and analytical evidence, we conjecture
that at small positive energies the phase space of our model is completely
chaotic except for a single region of complete integrability with a smooth
sharp boundary showing no KAM-type structures of any kind. This results in
the cleanest mixed phase space structure possible, in which motions in the
integrable region and in the chaotic region are clearly separated and
independent of one another. For certain system parameters, this mixed phase
space structure can be tuned to make either of the two components disappear,
leaving a completely integrable or completely chaotic phase space. For other
values of the system parameters, additional structures appear, such as
KAM-like elliptic islands, and one parameter families of parabolic periodic
orbits embedded in the chaotic sea. The latter are analogous to bouncing
ball orbits seen in the stadium billiard. The analytical part of our study
proceeds from a geometric description of the dynamics, and shows it to be
equivalent to a linked twist map on the union of two intersecting disks.
\end{abstract}

\maketitle

\section{Introduction}

\label{intro}

The one-dimensional free particle and the one-dimensional harmonic
oscillator are arguably the two simplest quantum mechanical systems that
exist. Nonetheless, and in spite of the intrinsic simplicity of such systems
when treated in isolation, the problem of an essentially free particle
interacting locally with one or more oscillators is difficult and important
to a large class of physical systems. It arises in a variety of contexts,
ranging from fundamental studies aimed at understanding the emergence of
dissipation in Hamiltonian systems \cite{Debievre}, to electron-phonon
interactions in solids \cite{Holstein,electronphonon}, and, more recently,
to basic issues associated with the phenomena of quantum decoherence. In the
condensed matter literature, in particular, a great deal of theoretical work
has focused on the nature of electron-phonon interactions, and the rich
variety of behavior that occurs, such as the emergence of ``polaronic''
quasi-particles. In one version of the well-known Holstein Molecular Crystal
Model, a tight-binding electron in a crystal moves between different unit
cells, in each of which it interacts with a local oscillator \cite{Holstein}%
. Even in the simplest ``spin-boson'' form of this problem, in which the
particle can be conceived as moving between just two sites, and in which it
interacts with a single collective oscillator, the problem is not exactly
soluble, and has been the subject of intense investigations regarding the
appropriateness of various semi-classical approximations \cite{SpinBoson}.

In this paper we show that the situation can be just as complex in
completely classical versions of the problem. We study here the surprisingly
rich classical dynamics of what is perhaps the simplest Hamiltonian model
that one can think of that incorporates the essential features of this local
free particle-oscillator interaction \cite{oscillators}. Specifically, we
consider a single classical particle of mass $m,$ position $x,$ and momentum 
$p_{0}$ that moves on a ring, the circumference of which is divided into two
sections. On one section the particle moves freely, on the other it
interacts with a single oscillator of mass $M$ and frequency $\omega$. The
associated Hamiltonian we write in the form%
\begin{equation}
H=\frac{p_{0}^{2}}{2m}+\frac{P^{2}}{2M}+\frac{1}{2}M\omega^{2}X^{2}-F_{0}X%
\rho(x),   \label{H1}
\end{equation}
where $P$ is the oscillator momentum, and $F_{0}$ describes the strength of
the interaction, which is linear in the oscillator coordinate $X$, but not
in the particle position $x$. In the Hamiltonian (\ref{H1}), $\rho(x)$ is a
form factor localized around $x=0$ that describes the range of the
interaction. We take $\rho$ to equal unity throughout the interaction region 
$\left| x\right| \leq\hat{\sigma}$, and to vanish on the section of the ring 
$\hat{\sigma} + \hat{L}/2>\left| x\right| >\hat{\sigma}$ lying outside this
range. With this choice, except at those moments of the evolution when the
particle arrives at $x=\pm\hat{\sigma}$, the particle and the oscillator are
effectively uncoupled, and evolve independently. At ``impacts'', i.e., when
the particle arrives at the edges of the interaction region, the particle
receives an impulsive kick from the oscillator that conserves the total
energy of the system. After each such kick, the particle again moves as a
free particle either inside or outside of the interaction region. It is this
essentially uncoupled evolution of the two subsystems between impulsive
kicks of the particle at $x=\pm\hat{\sigma}$ that makes it possible to
numerically and analytically track the dynamics.

\begin{figure}[t]
\centering \includegraphics[width=7in]{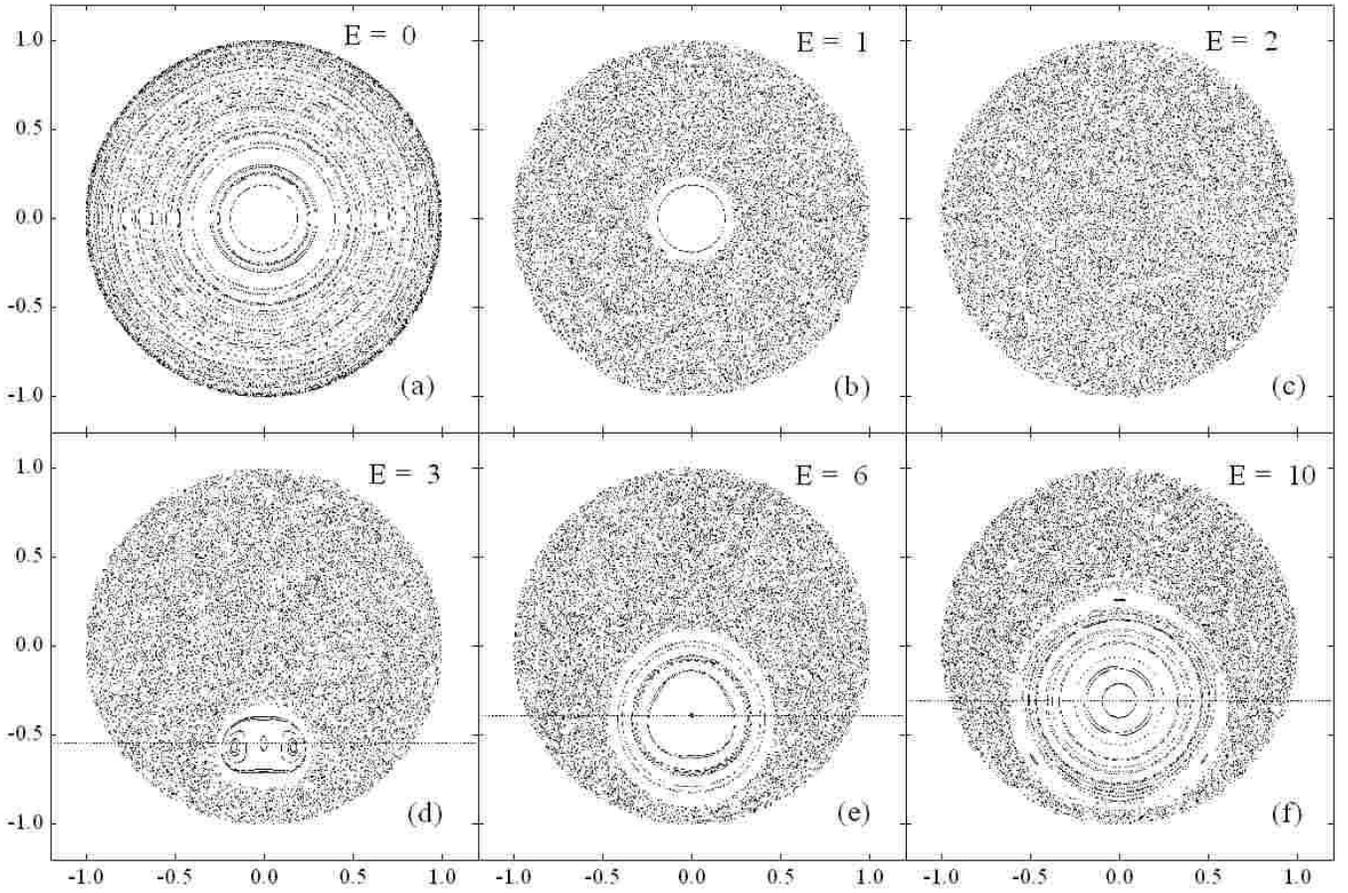} 
\caption{{Oscillator phase space diagrams for a system with $\protect\alpha=
2$ and $L=4.74$, and varying values of the energy $E=0,1,2,3,6,$ and $10$,
as shown (See Sec.\ \protect\ref{reduce} for the definition of the reduced
parameters used). The vertical axis in each figure is the scaled oscillator
coordinate $\protect\zeta- d$, and the horizontal axis the reduced
oscillator momentum $\protect\eta$ defined in Eq. \protect\ref{ZetaEta}. }}
\label{Eseries}
\end{figure}

\begin{figure}[t]
\centering \includegraphics[width=7in]{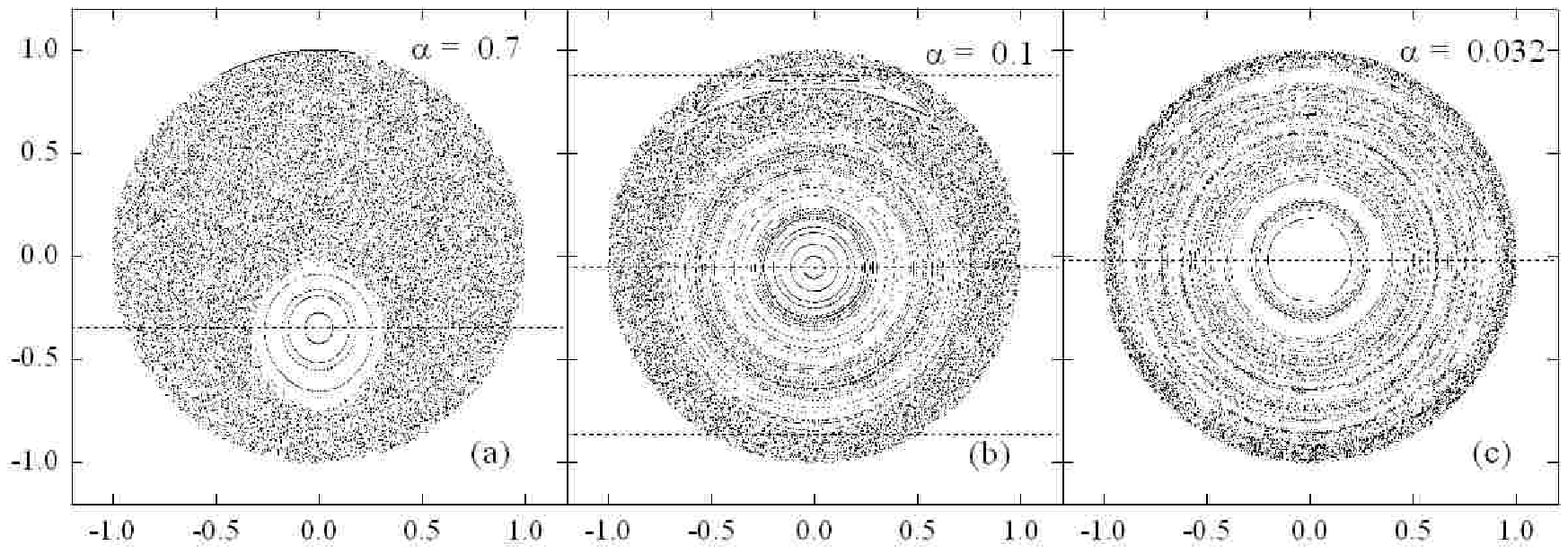} 
\caption{{Oscillator phase space diagrams for a system with $E=0.85$ and $L=2
$, and values of the coupling strength $\protect\alpha=0.7, 0.1$, and $0.032$
(See Sec.\ \protect\ref{reduce} for the definition of the reduced parameters
used). The vertical axis in each figure is the scaled oscillator coordinate $%
\protect\zeta- d$, and the horizontal axis the reduced oscillator momentum $%
\protect\eta$ defined in Eq. \protect\ref{ZetaEta}. The prominent elliptic
island in these figures is what is referred to in the text as Void II. In
the last two figures other secondary KAM structures appear at the edges of
the Void. }}
\label{AlphaSeries}
\end{figure}

\begin{figure}[t]
\centering \includegraphics[width=7in]{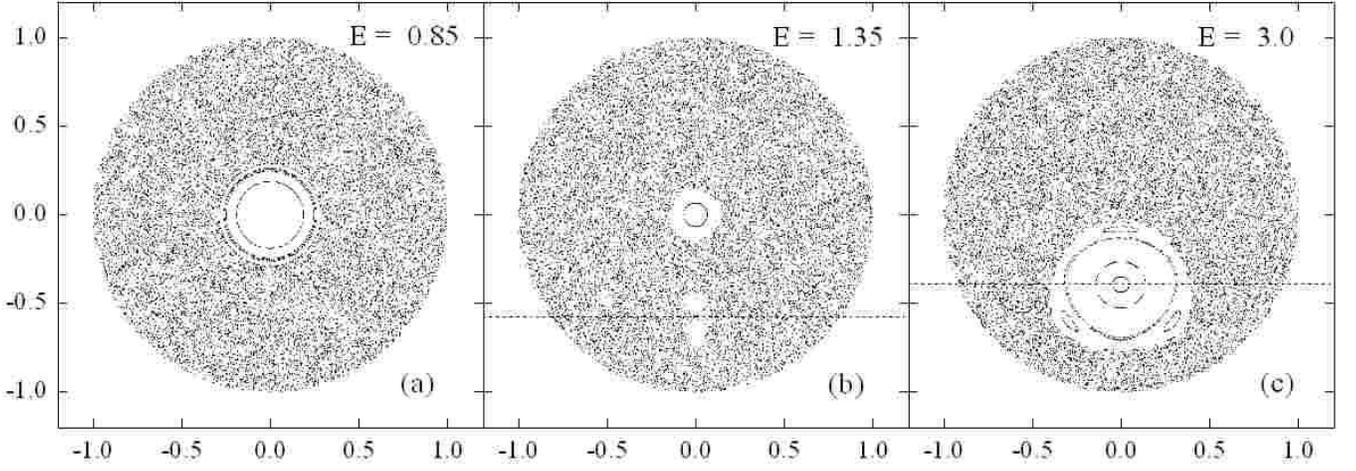} 
\caption{{Oscillator phase space diagrams for a system with $\protect\alpha=
2$ and $L=2.105$, and energy $E = 0.85, 1.35$, and $3$. The vertical axis in
each figure is the scaled oscillator coordinate $\protect\zeta- d$, and the
horizontal axis the reduced oscillator momentum $\protect\eta$ defined in
Eq. \protect\ref{ZetaEta}. } }
\label{EseriesLow}
\end{figure}

\begin{figure}[t]
\centering \includegraphics[width=7in]{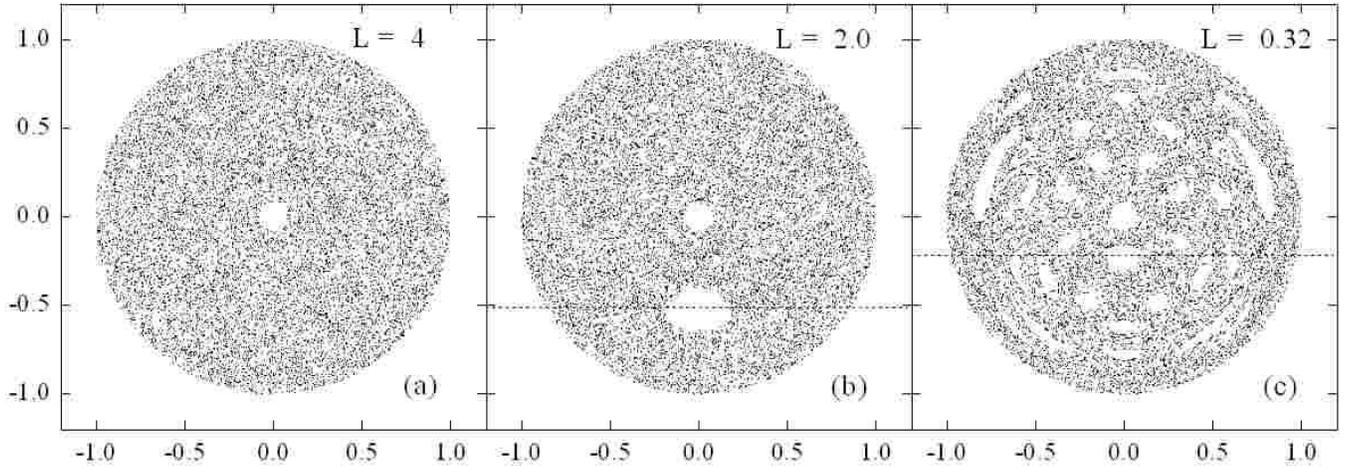} 
\caption{{Oscillator phase space diagrams for a system with $E=1.6$ and $%
\protect\alpha=2$, and values of the length $L=4, 2,$ and $0.32$, of the
non-interacting region as shown (See Sec.\ \protect\ref{reduce} for the
definition of the reduced parameters used). The vertical axis in each figure
is the scaled oscillator coordinate $\protect\zeta- d$, and the horizontal
axis the reduced oscillator momentum $\protect\eta$ defined in Eq. \protect
\ref{ZetaEta}. At the center of each of these figures is a small Void I. As $%
L$ is decreased, Void II appears and then ascends toward the impenetrable
Void I. The collision between these two elliptic islands appears to generate
considerable structure.}}
\label{Lseries}
\end{figure}

Despite its apparent simplicity, and as can be seen in the numerically
determined phase plots appearing in Figs.\ \ref{Eseries}-\ref{Lseries}, the
model has a striking combination of features that are not often seen
together in a single closed Hamiltonian system. First, for any given system
parameters, the phase space has a very simply described, completely
integrable region of controllable size for all energies ranging from that of
the ground state, $\hat{E}_{g}=-F_{0}^{2}/2M\omega^{2}$, up to a critical
positive energy $\hat{E}_{c}=|\hat{E}_{g}| $ (See Fig. \ref{Eseries} (a) and
Fig.\ \ref{Eseries} (b)). For small positive energies the region of phase
space outside of this region, which we refer to as Void I, appears in our
numerical calculations to be fully chaotic, with no secondary KAM
structures, even very close to the edge of the Void (See, e.g., Fig.\ \ref%
{Eseries}(b)). This then makes for the simplest possible mixed phase space
structure, in which motions in the completely integrable region and in the
chaotic region are clearly separated and independent of one another. We are
unaware of other Hamiltonian systems which exhibit this clean separation,
although it does appear in a piecewise linear symplectic map on the torus
that was explicitly constructed to exhibit this property \cite{MP}.

As a second striking feature of the model, we identify one-parameter
families of marginally unstable periodic orbits, that appear as circular
arcs in the oscillator phase space (See Fig. \ref{ArcFig}), and are similar
to the so-called bouncing ball orbits that arise in the stadium billiard.

We furthermore show that for fixed, suitably-chosen system parameters, the
statistical properties of the dynamics on the different energy surfaces can
vary greatly. For some values of the energy, the dynamics is completely
integrable, whereas for others it is fully chaotic or displays a mixed,
KAM-type behavior. Among the rich variety of structures that arise, we
identify and locate the central fixed point of a KAM-type elliptic island
that we refer to as Void II. In some cases Void II appears alone, as in
Fig.\ \ref{Eseries}(d)-(f) and Fig.\ \ref{AlphaSeries}, while in others Void
I and Void II both appear, as in Fig.\ \ref{Lseries}(b) and \ref{Lseries}%
(c). Finally, in the limit of small coupling strengths $F_{0}$, the phase
space displays typical KAM structures of the type that generally occur when
a completely integrable motion is subject to a small nonlinear perturbation,
as in Fig.\ \ref{AlphaSeries}. This, however, by no means exhausts the
variety of structures that seem to appear (See Fig. \ref{pigface}).

The presence of chaos in our coupled particle-oscillator model can be
understood intuitively as follows. A trajectory of the combined system will
tend to be unstable whenever the time the particle takes between two impacts
at $x=\pm\hat{\sigma}$ is long compared to the oscillator period. Indeed,
when the particle goes slowly, a small change in its velocity at one impact
will lead to a large change in the relatively fast moving oscillator
coordinate at the next impact. As a result, the height of the potential
barrier the particle meets at that moment becomes highly unpredictable, and
this is the source of the instability. That this simple picture is in
agreement with observed behavior can be seen in Fig.\ \ref{EseriesLow}, in
which it is clear for a given $F_{0}$ and $\hat{L}$ (represented in that
figure by the dimensionless parameters $\alpha$ and $L$ introduced in the
next section), that the fraction of phase space outside Void I that is
chaotic tends to increase as the total energy of the system (and thus the
particle speed) decrease. Similarly, increasing $\hat{L}$ tends to increase
the chaotic fraction of phase space outside of Void I, since for larger $%
\hat{L}$ the particle spends more time between succesive visits at 
$x=\pm\hat{\sigma}$ whenever it is outside the interaction region. A more 
precise and quantitative analysis of the mechanism leading to chaos in this 
model will be given in Sec. \ref{geo}, where we will show that the dynamics 
can be analyzed in terms of a discontinuous linked twist map on the union of 
two intersecting disks. This is a generalization of the linked twist maps on 
the torus introduced and studied in \cite{TwistMap}.

The rest of the paper is laid out as follows. In the next section, we
simplify the Hamiltonian (\ref{H1}) by reducing the six system parameters
with which it is associated down to two. Following this reduction, we
present phase-space plots of the results of a numerical integration of the
equations of motion for the system. In Sec.\ \ref{geo}, we develop a
geometric description of the dynamics that makes it relatively
straightforward to explain the main features seen in numerical studies,
including the emergence of chaos at positive energies (Sec.\ \ref{chaos}),
the existence of islands of regular motion of two different generic types,
which we refer to as Void I and Void II (studied in Secs.\ \ref{confined}
and \ref{voidTwo}, respectively), and the presence of arcs of marginally
stable periodic orbits that appear as sets of zero measure in the chaotic
portions of our phase space diagrams (Sec.\ \ref{confined}). In addition, we
analytically demonstrate the existence of unstable isolated periodic orbits
of arbitrarily large period (Sec.\ \ref{voidTwo}). In the last section, we
summarize our results, and comment on their ramifications.

\section{Dimensional Reduction of the Hamiltonian}

\label{reduce}
 
The system described by the Hamiltonian $H(x,p_{0},X,P)$ as
given in (\ref{H1}) depends upon six system parameters: the two masses $m$
and $M,$ the oscillator frequency $\omega,$ and the coupling strength $F_{0},
$ as well as the widths $2\hat{\sigma}$ and $\hat{L}$ of the interacting and
non-interacting sections of the ring on which the particle moves. To reduce
the number of inessential parameters in the system, we now introduce a
dimensionless time $\tau=\omega t,$ and dimensionless variables and momenta%
\begin{equation}
\tilde{p}=\frac{p_{0}}{\sqrt{m\hbar\omega}}\qquad\tilde{q}=\sqrt{\frac {%
m\omega}{\hbar}}x,\qquad\tilde{\Phi}=\sqrt{\frac{M\omega}{\hbar}}X,\qquad%
\tilde{\Pi}=\frac{P}{\sqrt{M\hbar\omega}}   \label{VariablesTilde}
\end{equation}
where $\hbar$ is an arbitrary constant having units of action that plays no
part in the subsequent dynamics. The new variables obey the equations of
motion 
\begin{align}
\frac{d\tilde{q}}{d\tau} & =\tilde{p}\qquad\frac{d\tilde{p}}{d\tau}=\tilde{%
\alpha}\tilde{\Phi}\frac{d\tilde{\chi}}{d\tilde{q}}\left( \tilde {q}\right) 
\notag \\
\frac{d\tilde{\Phi}}{d\tau} & =\tilde{\Pi}\qquad\frac{d\tilde{\Pi}}{d\tau }=-%
\tilde{\Phi}+\tilde{\alpha}\tilde{\chi} (\tilde{q})   \label{EOMTilde}
\end{align}
which are the canonical equations generated by a transformed Hamiltonian $%
\tilde{H}_{\tilde{\alpha}}=H/\hbar\omega$, where 
\begin{equation}
\tilde{H}_{\tilde{\alpha}}=\frac{1}{2}\left( \tilde{p}^{2}+\tilde{\Pi}^{2}+%
\tilde{\Phi}^{2}\right) -\tilde{\alpha}\tilde{\Phi}\tilde{\chi}(\tilde {q}), 
\label{Htilde}
\end{equation}
$\tilde{\alpha}=\sqrt{F_{0}^{2}/M\hbar\omega^{3}},$ and the function $\tilde{%
\chi}(\tilde{q})=\rho(\tilde{q}\sqrt{\hbar/m\omega})$ vanishes outside the
interaction region extending between $\tilde{q}=\pm\sigma\equiv\pm \hat{%
\sigma}\left( m\omega/\hbar\right) ^{1/2}.$ Explicitly, we write 
\begin{equation}
\tilde{\chi}\left( \tilde{q}\right) \equiv\chi\left( \tilde{q}/\sigma\right)
=\theta\left( \frac{\tilde{q}}{\sigma}+1\right) -\theta\left( \frac{\tilde{q}%
}{\sigma}-1\right)   \label{lambda}
\end{equation}
where $\theta\left( x\right) $ is the Heaviside step function. This form for
the function $\chi$ allows an additional simplification through the scale
transformation 
\begin{equation}
q=\tilde{q}/\sigma\qquad p=\tilde{p}/\sigma\qquad\Phi=\tilde{\Phi}%
/\sigma\qquad\Pi=\tilde{\Pi}/\sigma.   \label{Finalpq}
\end{equation}

With a suitable redefinition of the coupling constant 
$\tilde{\alpha}=\alpha\sigma$, we obtain the following one parameter family 
\begin{equation}
H_{\alpha}=\frac{1}{2}\left( p^{2}+\Pi^{2}+\Phi^{2}\right) -\alpha\Phi
\chi(q)   \label{Halpha}
\end{equation}
of reduced Hamiltonians $H_{\alpha}=\tilde{H}_{\tilde{\alpha}}\sigma^{-2}$
describing the system, where now the interaction region is associated with
the fixed interval $q\in\left[ -1,1\right] $. Note that in this form the
dimensionless coupling constant 
\begin{equation}
\alpha= \sqrt{|\hat E_{g}|/\frac12{m\hat\sigma^{2}\omega^{2}}}
\end{equation}
involves the ratio of the (original) ground state energy of the system to
the kinetic energy of a particle that crosses the interaction region in one
oscillator period.

We have thus reduced the number of system parameters down to the single
explicit parameter $\alpha$ describing the coupling strength, and one
additional parameter $L=\hat{L}/\hat{\sigma}$ associated with the total
range $2+L$ of the particle coordinate $q$. We note in passing that with
this choice for the function $\chi\left( q\right) ,$ the coupling parameter
is equal to the equilibrium value $\Phi_{\text{eq}}^{\text{in}}=\alpha$ of
the oscillator coordinate when the particle is in the interaction region $%
q\in\left[ -1,1\right] ,$ and that the ground state, which now has rescaled
energy $E_{g}=-\alpha^{2}/2,$ occurs when the particle and the oscillator
are both at rest, with the particle in the interaction region.

The equations of motion corresponding to (\ref{Halpha}) 
\begin{align}
\dot{q} & =p\qquad\qquad\dot{p}=\alpha\Phi\left[ \delta\left( q+1\right)
-\delta\left( q-1\right) \right]  \notag \\
\dot{\Phi} & =\Pi\qquad\qquad\dot{\Pi}=-\Phi+\alpha\chi\left( q\right) , 
\label{EOMfinal}
\end{align}
with dots denoting derivatives with respect to $\tau$, show that the
particle feels an impulsive force only when it reaches the edges of the
interaction region, but otherwise travels as a free particle. The particle's
momentum undergoes discontinuous changes at these moments, but its position
remains a continuous function of time. The impulse imparted to the particle
at $q=\pm1$ is readily computed from the oscillator displacement using only
conservation of total energy. When the particle enters the interaction
region, the oscillator experiences an interaction force of finite magnitude 
$\alpha$ that suddenly shifts the equilibrium position about which it
oscillates from $\Phi_{\text{eq}}^{\text{out}}=0$ to $\Phi_{\text{eq}}^
{\text{in}}=\alpha$, but the amplitude and momentum of the oscillator remain
continuous functions of time. Thus, when the particle is in the interaction
region, the oscillator phase point $\left( \Phi\left( \tau\right) ,\Pi\left(
\tau\right) \right) $ rotates at unit angular speed about the equilibrium
state $\left( \Phi,\Pi\right) =\left( \alpha,0\right) ,$ and when the
particle is outside the interaction region, a similar free rotation takes
place about the origin of the oscillator phase plane. As mentioned in the
introduction, it is this essentially uncoupled evolution of the two
subsystems between impulsive kicks of the particle at $q=\pm1$ that makes it
straightforward to numerically and analytically track the resulting dynamics.

\begin{figure}[t]
\centering \includegraphics[height=2in]{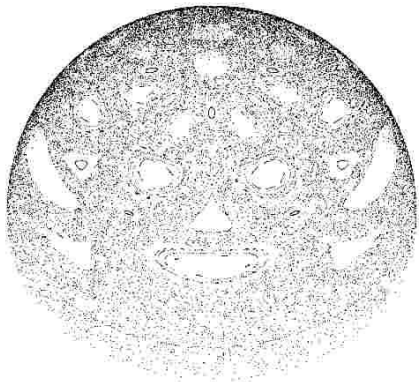} 
\caption{{Visualization of the Poincar\'e section at $x = 0$ of the energy
surface for a particle-oscillator system with $E=1.6$, $\protect\alpha= 2$,
and $L=0.32$ (See Sec.\ \protect\ref{reduce} for the definition of the
reduced parameters used). We display phase points generated, as described in
the text, for 100 initial conditions randomly chosen on the energy surface.
The same data appear in Fig.\ \protect\ref{Lseries}(c). This representation
of the energy surface is as seen by an observer located outside the energy
surface, upside down in the $\Pi$-$p$ plane. }}
\label{pigface}
\end{figure}

As in any system with two degrees of freedom, the energy surfaces associated
with (\ref{Halpha}) are three dimensional. Their two-dimensional sections at 
$q=0$ are ellipsoids centered at $\left( \Phi,\Pi,p\right) =\left(
\alpha,0,0\right) .$ In Fig.\ \ref{pigface} we display phase points recorded
on the Poincar\'e section at $q=0$ and $p>0$ arising from the numerically
determined evolution of 100 initial conditions randomly chosen on such an
energy surface and evolved for a time corresponding to 500 passages through
the section. Note that the Hamiltonian (\ref{Halpha}) is invariant with
respect to the parity operation of the particle. Thus, if $\left( \Phi\left(
t\right) ,\Pi\left( t\right) ,p\left( t\right) ,q\left( t\right) \right) $
is a solution to the equations of motion, so is $\left( \Phi\left( t\right)
,\Pi\left( t\right) ,-p\left( t\right) ,-q\left( t\right) \right) $, and for
symmetrized pairs of randomly chosen initial conditions, the corresponding
set of phase points recorded at $-p$ will be the same as at $p,$ i.e., the
back half of the ellipsoid, if displayed, would be a mirror image of the
front half. This symmetry, together with time reversal invariance explains
the additional $\Pi\rightarrow-\Pi$ symmetry that our phase plots exhibit.

As a result it therefore suffices to represent the evolution by simply
recording the \emph{oscillator phase point} $\left( \Phi\left( t\right)
,\pi\left( t\right) \right) $ each time that $q=0$ and $p$ is positive. Such
a representation is obtained by numerically computing the return map for
this Poincar\'e section, and has been used in Figs.\ \ref{Eseries}-\ref%
{Lseries} and \ref{ArcFig}.

\section{Geometric Description of the Dynamics}

\label{geo}

\begin{figure}[t]
\centering \includegraphics[height=2.5in]{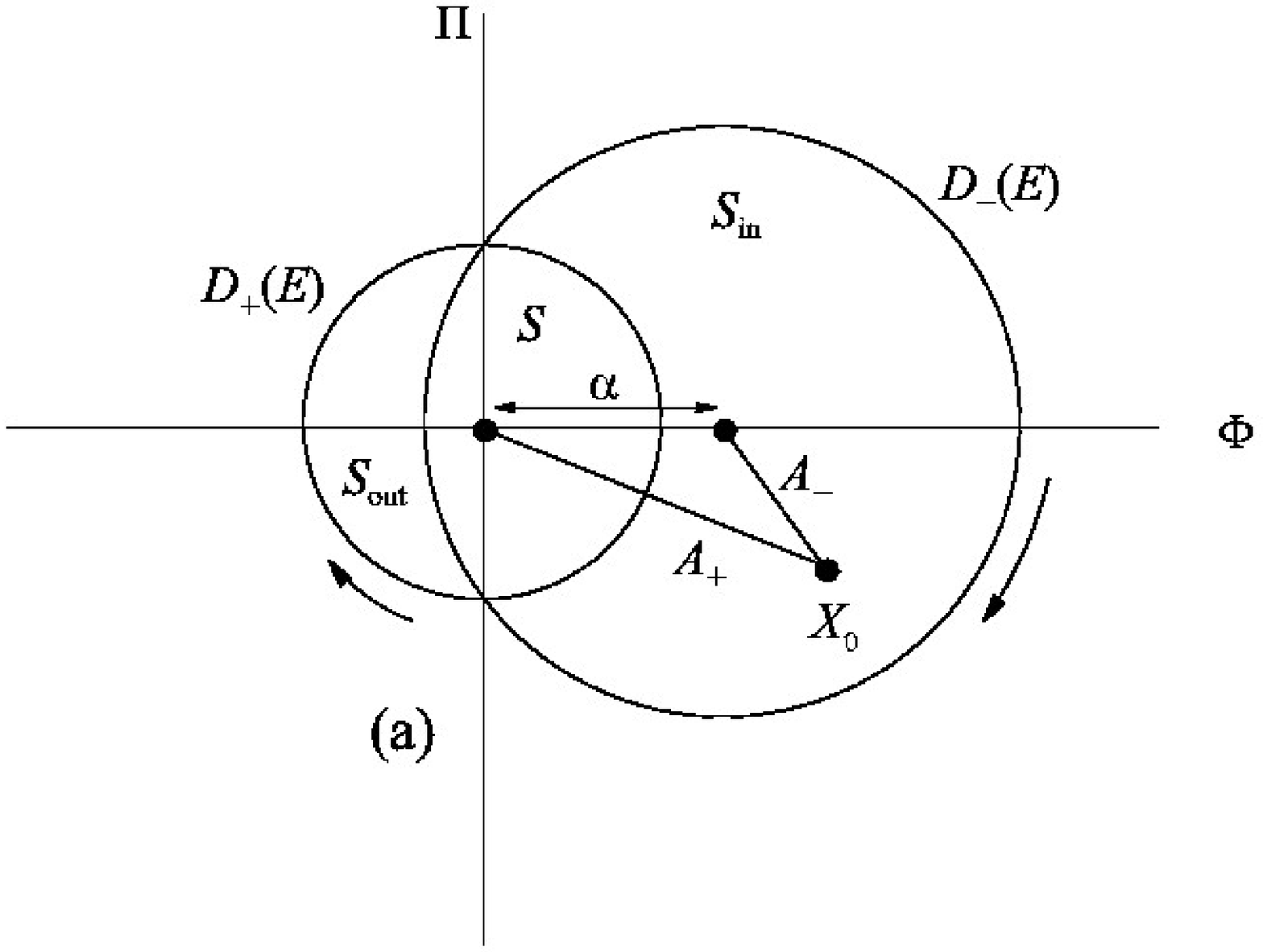} %
\includegraphics[height=2.5in]{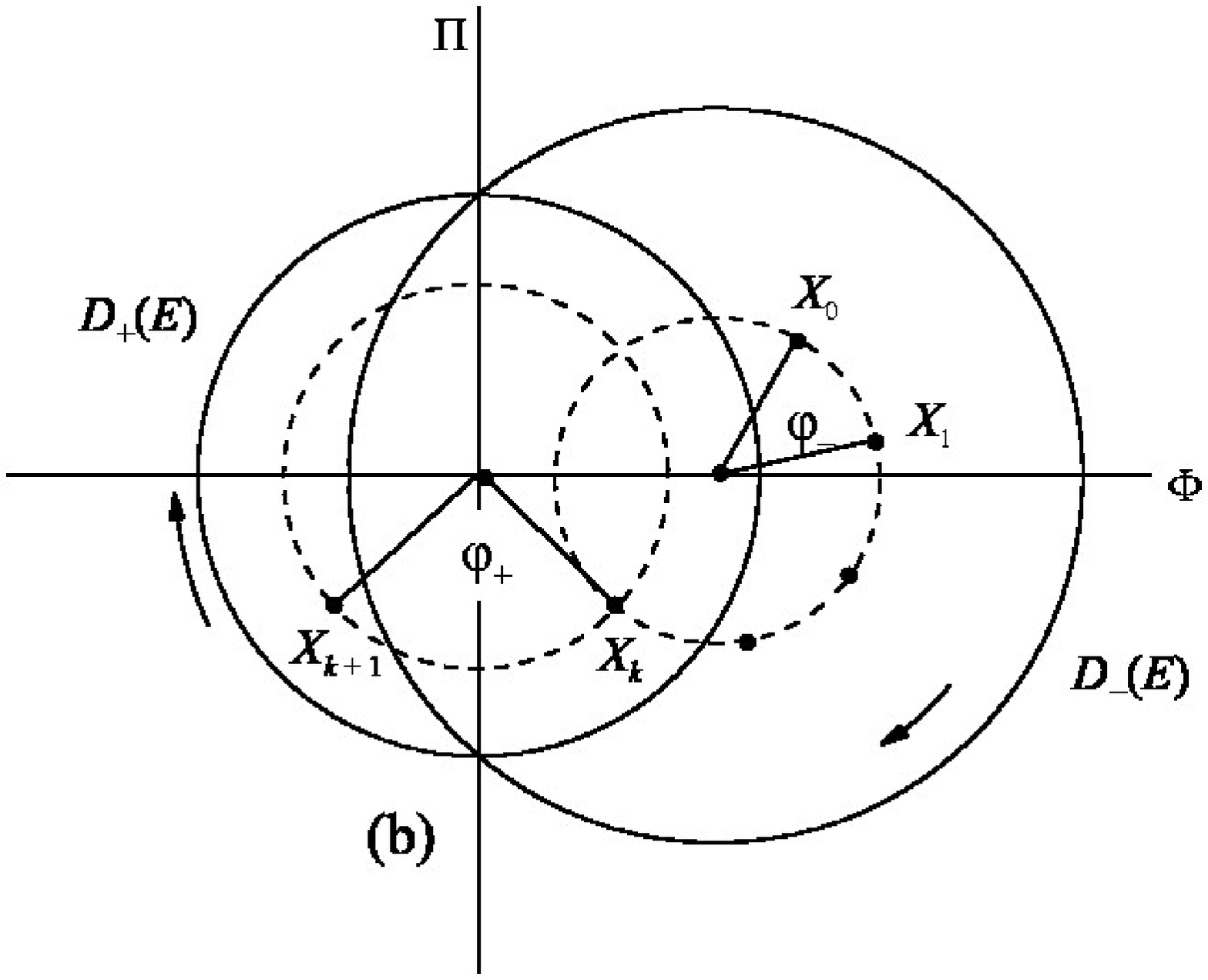} 
\caption{{The oscillator phase plane showing the disks $D_{+}(E)$ and $%
D_{-}(E)$ for (a) $E < \protect\alpha^{2}/2$, and (b) for $E > \protect\alpha%
^{2}/2$. On these figures the horizontal axis corresponds to the oscillator
coordinate $\Phi$ and the vertical axis to the oscillator momentum $\Pi$.}}
\label{D1}
\end{figure}

Although the dynamics of the system recorded at $q=0$ shows most
clearly the inherent symmetries of the system, to actually explain the
features that appear in our phase plots, it is obviously necessary to
consider those times at which the particle reaches the edges of the
interaction region at $q=\pm1$. At fixed energy $E$, and with the position $%
q=\pm1$ of the particle determined up to a sign, the state of the system at
these moments is also conveniently represented as a point in the $\Phi-\Pi$
oscillator phase plane. In general, at a given $E>0$, the available phase
space for the oscillator when the particle is in the interaction zone is a
disk $D_{-}(E)$ of radius $\sqrt{2E+\alpha^{2}}$ centered at the point $%
(\alpha,0)$ in the $\left( \Phi,\Pi\right) $ plane, that obviously contains
the origin of the phase plane. Similarly, when the particle is outside the
interaction zone, the available phase space for the oscillator is a disk $%
D_{+}(E)$ of radius $\sqrt{2E}$ centered at the origin. Any point $\left(
\Phi,\Pi\right) $ in the set $S_{\text{in}}\left( E\right)
=D_{-}(E)\setminus D_{+}(E),$ i.e., inside $D_{-}(E)$ but outside $D_{+}(E),$
corresponds to a state with the particle definitely inside the interaction
region, and any point in $S_{\text{out}}\left( E\right) =D_{+}\left(
E\right) \setminus D_{-}\left( E\right) $ to a state with the particle
definitely outside the interaction region. Depending on the energy and the
coupling constant, two distinct situations can occur: either the disk $%
D_{+}\left( E\right) $ does not contain the center of $D_{-}\left( E\right) $%
, which occurs for $E<\alpha^{2}/2,$ and is depicted in Fig. \ref{D1}(a), or
it does contain the center of $D_{-}\left( E\right) $, which occurs when $%
E>\alpha^{2}/2$ and is depicted in Fig. \ref{D1}(b). It should be obvious
from the geometry of both figures that the circles which form the edges of
the two disks always intersect on the $\Pi$ axis. In the following, given a
point $X=(\Phi,\Pi)$, we denote by $A_{\pm}$ its distance from the center of 
$D_{\pm}(E)$. Physically, $\frac{1}{2}A_{+}^{2}=\frac{1}{2}\left(
\Phi^{2}+\Pi^{2}\right) $ is the uncoupled oscillator energy.

We can now give a simple qualitative description of the motion of the system
in geometric terms (see Fig. \ref{D1}(b)). Suppose that, for fixed $E>0,$
the system is such that at $t=0$ the particle is on the left edge of the
interaction zone and moving inward, i.e., $q(0)=-1$ and $\dot{q}(0)>0,$ so
that $X_{0}=(\Phi_{0},\Pi_{0})\in D_{-}(E)$. As the particle now crosses the
interaction region to $q=1$, the oscillator phase point simply rotates on a
circular arc around the point $(\alpha,0)$ through an angle that, in our
dimensionless units, can be written 
\begin{equation}
\varphi_{-}(A_{-})=\frac{2}{\dot{q}(0)}=\frac{2}{\sqrt{2E+%
\alpha^{2}-A_{-}^{2}}},   \label{phim}
\end{equation}
where the second form follows through energy conservation. Note that the
rotation angle depends monotonically on the radius of the circle on which
the phase point moves, so that two nearby phase points at different radii
will rotate through different angles. In this way a \emph{shear} $T_{-}$ is
induced on the disk $D_{-}\left( E\right) .$ Since the angle 
$\varphi_{-}(A_{-})$ diverges as $A_{-}$ approaches the radius of $D_{-}(E)$,
the strength of the shear diverges near the edge, and the dynamics becomes
increasingly sensitive to small changes in $A_{-}$.

If the new phase point $X_{1}$ so obtained lies in $S_{\text{in}}\left(
E\right) ,$ the particle encounters a potential energy barrier at $q=1$ that
is greater than its kinetic energy. Consequently, it reflects from the
barrier, reversing its motion to cross the interaction region back to $q=-1,$
during which time the oscillator phase point continues its rotation about
the center of $D_{-}(E)$, through the same angle $\varphi_{-}(A_{-})$. The
process then repeats itself, generating a sequence of oscillator phase
points $X_{1},X_{2,},\ldots$ separated by equal angular displacements 
$\varphi _{-}(A_{-})$ until, for some $k,$ the phase point $X_{k}$ falls
inside $S\left( E\right) .$

At such a time, the particle has sufficient kinetic energy to overcome the
barrier it encounters, and passes out of the interaction region with a new
kinetic energy equal to $E-\frac{1}{2}A_{+}^{2},$ where $A_{+}$ is the
distance between the point $X_{k}$ and the origin (the center of the disk 
$D_{+}\left( E\right) $). The particle then travels through a distance $L$
around the outer section of the ring, and arrives again at $q=\pm1,$ while
the oscillator phase point rotates on the \emph{second} disk $D_{+}\left(
E\right) $ through an angle 
\begin{equation}
\varphi_{+}(A_{+})=\frac{L}{\sqrt{2E-A_{+}^{2}}}   \label{phip}
\end{equation}
about the origin of the phase plane. Thus, a \emph{different} shear $T_{+}$
is induced on the disk $D_{+}\left( E\right) $. The new phase point $X_{k+1}$
so obtained is depicted in Fig.\ \ref{D1}(b). Obviously this way of
depicting the evolution can be repeated indefinitely, and provides an
efficient way to analyze and explain various features of the dynamics.

\section{The Emergence of Chaos}

\label{chaos}

For example, the description given in the last section makes
it clear that the mechanism underlying the appearance of chaos in this
system is the existence of the two non-aligned shears $T_{+}$ and $T_{-}$.
This is a well-known phenomenon. For example, on a torus 
$x,y\in\left[ 0,1\right]$, successive application of the two shears%
\begin{equation*}
\left( 
\begin{array}{c}
x \\ 
y%
\end{array}
\right) \rightarrow\left( 
\begin{array}{c}
x+ay \\ 
y%
\end{array}
\right) \qquad\text{and}\qquad\left( 
\begin{array}{c}
x \\ 
y%
\end{array}
\right) \rightarrow\left( 
\begin{array}{c}
x \\ 
y+bx%
\end{array}
\right) \qquad 
\end{equation*}
yields a hyperbolic map%
\begin{equation*}
\left( 
\begin{array}{c}
x \\ 
y%
\end{array}
\right) \rightarrow\left( 
\begin{array}{cc}
1 & a \\ 
b & 1+ab%
\end{array}
\right) \left( 
\begin{array}{c}
x \\ 
y%
\end{array}
\right) 
\end{equation*}
and leads to chaotic dynamics if $\left| 2+ab\right| >2.$ Locally, the two
shears $T_{+}$ and $T_{-}$ have exactly this structure, but in polar
coordinates and with the role of $a$ and $b$ played by $\varphi_{+}^{\prime
}(A_{+})$ and $\varphi_{-}^{\prime}(A_{-})$. The divergence of $\varphi_{\pm
}(A_{\pm})$ at the edges of the disks thus provides a clear mechanism for
the emergence of chaos in certain regions of phase space. In fact, the
dynamical system defined on the two disks described in the last section is a
generalization of what is referred to in the literature as a linked twist
map. Some simple examples of such maps (on a torus rather than on a union of
disks) have rigourously been shown to exhibit ergodicity and chaotic
behavior \cite{TwistMap}. The discontinuity of the functions $\varphi_{\pm}$
at the edges of the discs and the fact that the two shears are not
transverse everywhere in the intersection region would make a completely
rigorous analysis of the ergodic properties of our model considerably more
complicated.

\begin{figure}[t]
\centering \includegraphics[height=2.5in]{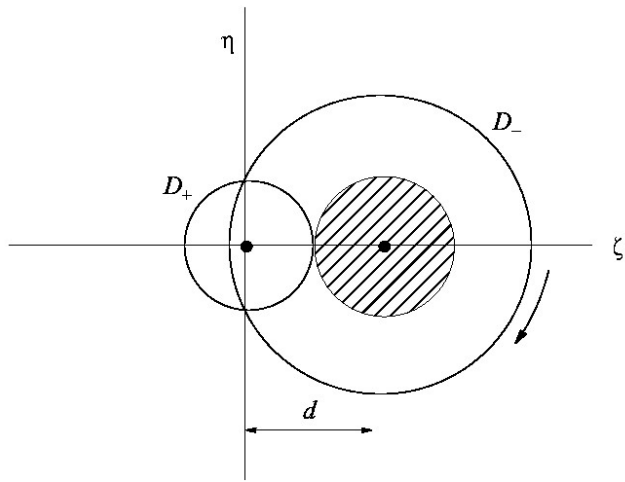} 
\caption{{The rescaled oscillator phase plane showing the disks $D_{+}$ and $%
D_{-}$. The shaded region indicates those initial conditions in $D_{-}$
which never enter the intersection $S$ of $D_{-}$ and $D_{+}$, and thus lead
to trajectories in which the particle remains in the interaction region.
These trajectories make up what we have referred to as Void I, with the
center of $D_{-}$ corresponding to the center of that elliptic island.}}
\label{VoidOneDisk}
\end{figure}

Note that in this essentially geometric description, the dynamics depends
only on three independent parameters $E,L$, and $\alpha$. It is helpful to
simplify the geometric nature of the description even further, by
reorganizing the parameters as follows. First, we re-express the coupling
strength of the system through the parameter%
\begin{equation}
d=\frac{\alpha}{\sqrt{2E+\alpha^{2}}}   \label{d-defined}
\end{equation}
and introduce rescaled oscillator variables 
\begin{equation}
\zeta=\frac{\Phi}{\sqrt{2E+\alpha^{2}}},\qquad\eta=\frac{\Pi}{\sqrt {%
2E+\alpha^{2}}}   \label{ZetaEta}
\end{equation}
which locate the oscillator phase point at radii 
\begin{equation}
r_{\pm}=\frac{A_{\pm}}{\sqrt{2E+\alpha^{2}}},\qquad0\leq r_{-}<1,
\qquad0\leq r_{+}<\sqrt{1-d^{2}}   \label{r-plus-minus}
\end{equation}
from the centers, respectively, of a disk $D_{-}\ $ of \emph{unit} radius
centered at $\left( d,0\right) ,$ and a disk $D_{+}$ of radius $\sqrt{1-d^{2}
}$ centered at the origin. With this choice, the dynamics now occurs on the
disks $D_{\pm},$ and the rotation angles $\varphi_{\pm}$ that the oscillator
phase points sweep through during one traversal of the interaction zone can
be simply written%
\begin{equation}
\varphi_{-}\left( r_{-}\right) =\varphi_{-}\left( A_{-}\right) =\frac{a_{-}}{%
\sqrt{1-r_{-}^{2}}},\qquad a_{-}=\frac{2}{\sqrt{2E+\alpha^{2}}} 
\label{phim(r)}
\end{equation}%
\begin{equation}
\varphi_{+}\left( r_{+}\right) =\varphi_{+}\left( A_{+}\right) =\frac{a_{+}}{%
\sqrt{1-d^{2}-r_{+}^{2}}},\qquad a_{+}=\frac{L}{\sqrt {2E+\alpha^{2}}}. 
\label{phip(r)}
\end{equation}
In this new description, we can thus take $d,$ $a_{+},$ and $a_{-}$ as the
three independent parameters describing the system$.$ Note that numerical
data in the oscillator phase plots appearing throughout this paper are
presented in terms of the scaled oscillator variables $\left(
\zeta-d,\eta\right) ,$ and thus always appear on a disk of unit radius
centered at the origin. This disk is essentially a shifted version of the
disk $D_{-}\ $defined above, but is rotated by 90 degrees, so that the
oscillator coordinate appears on the vertical, rather than the horizontal
axis. In what follows we will use both the scaled and unscaled variables as
is appropriate to the discussion at hand.

\begin{figure}[t]
\centering \includegraphics[height=3in]{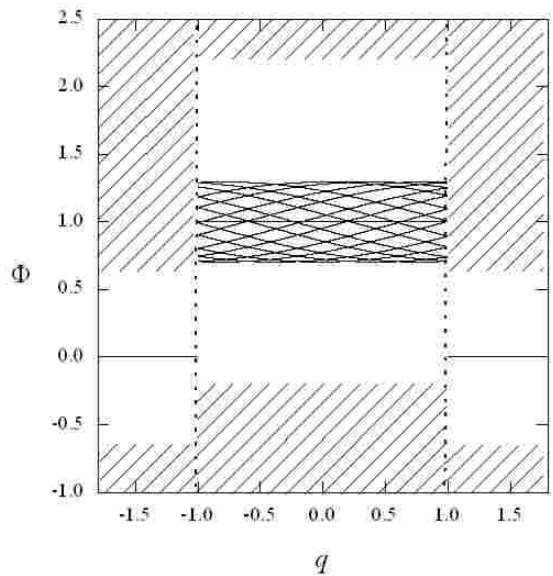} 
\caption{{A partial trajectory in the $q-\Phi$ plane for a motion in which
the particle remains confined to the interaction region. For this system $%
\protect\alpha= 1$ and $E =0.2$. Shaded portions indicate energetically
inaccessible regions of configuration space for a system with this energy,
and the region between the vertical dashed lines indicates the interaction
region as described in the text.} }
\label{PaulFig}
\end{figure}

\section{Motion Confined to the Interaction Region}

\label{confined}

In this section we analyze the motion of the system when
the particle never leaves the interaction region. In particular, we explain
the basic features of Void I, the region of complete integrability that
occurs at the center of the phase space plots in Figs.\ \ref{Eseries}(a) and
(b), and in Fig.\ \ref{Lseries}. In addition, we demonstrate and explain the
existence of arcs of parabolic fixed points embedded in the chaotic regions
of phase space.

The easiest case of confined motion to understand is the one in which the
total energy is negative, and the particle is \emph{energetically} confined
to the interaction region. For this situation the dynamics is totally
integrable, since the speed and kinetic energy of the particle are constants
of the motion, and the oscillator coordinate $\Phi(t)$ is never negative.
There are essentially three types of trajectories. The first type includes
those in which only the motion of the particle is excited, so $\Phi=\alpha$
and $\Pi=0$ are constant. Since the particle does not have enough kinetic
energy to overcome the barriers it encounters at $q=\pm1,$ it reflects at
each impact with the boundary. The trajectory is trivially periodic, with
period $4/\dot{q}$. In the second type of motion, the particle is at rest in
the interaction region and the oscillator performs simple harmonic motion of
amplitude $\Phi_{0}<\alpha$ about $\Phi=\alpha.$ Periodic orbits of this
type are also possible, of course, at any positive energy. Finally, there
are confined motions in which both of the ``modes'' described above are
excited. The trajectory is Lissajou-like, and the orbit is closed if the
oscillator period and the traversal time $2/\dot{q}$ are commensurate.
Otherwise it sweeps out a rectangle in the two-dimensional $\left(
q,\Phi\right) $ configuration space, with $q\in\left[ -1,1\right] ,$ and 
$\Phi\in\left[ \alpha-\Phi_{0},\alpha+\Phi_{0}\right] $, for some positive
amplitude $\Phi_{0}$ $<\alpha$ of oscillation.

When the total energy $E=0,$ possible motions with the particle confined to
the interaction region include all of the types described above. In
addition, oscillations of amplitude $\Phi_{0}=\alpha$ are now allowed; in
that case the particle is at rest somewhere in the interaction region. We
note that that the only other motions at $E=0$ are the unstable equilibrium
points corresponding to the particle at rest outside the interaction region
with the oscillator at rest as well.

\begin{figure}[t]
\centering \includegraphics[height=3in]{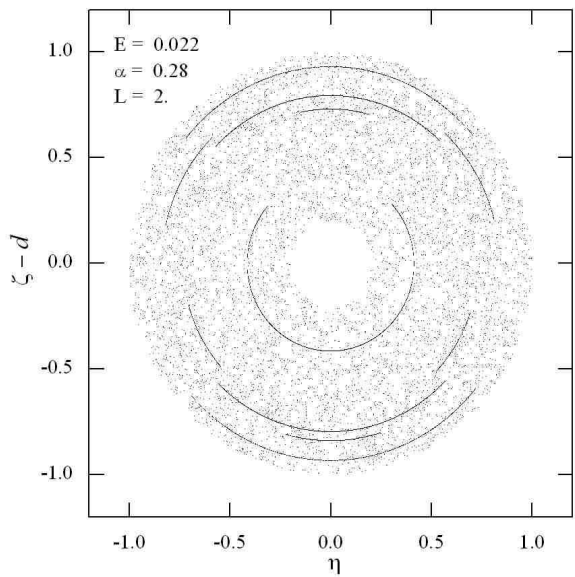} 
\caption{{Phase plot showing arcs of periodic orbits. In this figure one can
see a single arc of fixed points lying just outside of Void I, and, moving
out from the center, sets of arcs corresponding to orbits of period $k =
3,2,3,$ and $2$, corresponding, respectively, to $\ell=4,3,5$, and $5$.} }
\label{ArcFig}
\end{figure}

We now give a description of all orbits at positive energy for which the
particle never leaves the interaction zone. It is clear from Fig.\ \ref{D1}
(a), the scaled version of which appears in Fig.\ \ref{VoidOneDisk}, and the
geometric description of Sec. \ref{geo}, that for any energy $\alpha
^{2}/2>E>0,$ the disk $D_{+}$ does not contain the center of $D_{-}.$ Thus,
for this situation, all initial conditions with $r_{-}<R_{\text{I}},$
indicated by the shaded region in Fig.\ \ref{VoidOneDisk}, in which%
\begin{equation}
R_{\text{I}}= d-\sqrt{1-d^{2}},   \label{RadiusofVoidI}
\end{equation}
will give rise to trajectories for which the particle remains in the
interaction region. Indeed, since the circle of radius $r_{-}$ does not then
intersect $D_{+},$ successive rotations of such a point through 
$\varphi_{-}\left( r_{-}\right)$ can never lead to a point lying in $D_{+}.$
The elliptic regions of complete integrability at the center of Figs.\ \ref%
{Eseries}(a) and (b), and in Fig.\ \ref{Lseries}, which we have already
referred to as Void I, correspond precisely to trajectories of this kind. In
Fig. \ref{PaulFig} we show a typical periodic orbit of this type in the $%
q-\Phi$ plane. Shaded portions of that figure indicate classically forbidden
regions at this energy. From the equivalent point of view of a particle
moving in a two dimensional potential $V\left( q,\Phi\right) ,$ it is
interesting that the motion remains trapped within the interaction region
although there is no actual potential energy barrier preventing it from
leaving. Finally, we note that as $E$ is increased from zero at constant $%
\alpha$ the radius of Void I shrinks, as described by (\ref{RadiusofVoidI}),
until it finally disappears at the critical value $E=\alpha^{2}/2,$
(corresponding to $d=1/\sqrt{2}$), as in Fig.\ \ref{Eseries}(c). For $E\leq0$
Void I fills the entire oscillator phase space, as in Fig.\ \ref{Eseries}(a).

Aside from these trajectories that remain within Void I, there are at any
positive energy still other initial conditions that give rise to
trajectories in which the particle remains in the interaction region. As we
will show, however, such trajectories are then necessarily periodic. Indeed,
suppose $r_{-}>d-\sqrt{1-d^{2}}.$ Then the circle of radius $r_{-}$ \emph{%
does} necessarily intersect $D_{+},$ and successive rotations through $%
\varphi _{-}(r_{-})$ will take the orbit into $D_{+}$ whenever the rotation
angle $\varphi_{-}(r_{-})$ is an irrational multiple of $2\pi$. Thus for the
orbit to remain in $D_{-},$ it is necessary that the rotation angle $\varphi
_{-}(r_{-})=2\pi\ell/k$ be rational, and hence that the orbit be periodic.
Note that $k$ is then the period of the orbit (where it is understood that
the integers $k$ and $\ell$ are relatively prime). Furthermore, for such a $%
k-$periodic orbit to be possible, the phase points must somehow arrange to
miss the intersection with $D_{+}$ as they advance around $D_{-}\ $in
angular steps of $2\pi\ell/k.$ Clearly, for this to happen, the angle $%
\delta\left( r_{-}\right) $ subtended by the two intersection points of the
edge of the disk $D_{+}$ and the circle of radius $r_{-}$ on which the phase
point moves must be smaller than the angular separation between neighboring
orbit points, i.e., 
\begin{equation*}
\delta\left( r_{-}\right) < 2\pi/k. 
\end{equation*}
For fixed points $(k=1)$, this condition on the angle $\delta\left(
r_{-}\right) $ is automatically satisfied, but for periodic orbits with $%
k\geq2,$ the condition imposes a bound on $r_{-}$ that prevents, e.g.,
periodic orbits of this type occuring too close to the edge of $D_{-}.$

These arguments show that, in general, fixed points of the dynamics will
occur at the radii%
\begin{equation}
r_{\ell,-}=\sqrt{1-\left( \frac{a_{-}}{2\pi\ell}\right) ^{2}}\qquad
\ell>a_{-}/2\pi   \label{arcs}
\end{equation}
for which $\varphi_{-}\left( r_{\ell,-}\right) =2\pi\ell.$ Note that this
infinite sequence of values $r_{\ell,-}$ accumulates at $r_{-}=1,$ i.e., at
the edge of $D_{-}.$ It follows that any phase point in $S_{\text{in }%
}=D_{-}(E)\setminus D_{+}(E)$  with $r_{-}=r_{\ell,-}$ will be a fixed point
of the dynamics. The set of such points for a given value of $\ell$ is an
arc in the phase plane. In Fig. \ref{ArcFig}, such an arc of fixed points
with $\ell= 1$ appears near the edge of Void I.

Now suppose for some value of $k>1$ and $r_{-}$ we have $\varphi_{-}\left(
r_{-}\right) =2\pi\ell/k,$ and $\delta\left( r_{-}\right) <2\pi/k.$ In this
situation there will exist $k$ arcs of angular span $2\pi/k-\delta\left(
r_{-}\right) $ centered at $\left( d,0\right) ,$ each point of which is
associated with a period $k$ orbit. Several arcs associated with higher
order orbits of this type also appear in Fig.\ \ref{ArcFig}, and can
occasionally be seen in some of our other phase plots. If the phase points
were recorded at $q=\pm1,$ then each arc that appeared in such a diagram
would, by the arguments given above, lie outside of the intersection region
with $D_{-}$, which is at the bottom of each phase diagram in this paper.
However, because the phase points in these figures are recorded at $q=0$
when $p$ is positive, the angular position of each arc is rotated by an odd
multiple of $\varphi _{-}\left( r_{-}\right) /2$ from where it would be if
recorded at $q=\pm1.$ This is due to the rotation of the oscillator phase
point that occurs while the particle travels from the edge of the
interaction region back to the center, where the phase point is recorded.
Since this rotation is a function of the radius $r_{-}$, arcs of this kind
can generally appear at any orientation in the phase diagram.

On general arguments it is to be expected that periodic trajectories of this
kind are parabolic. Indeed, the stability matrix associated with a phase
point lying on such an arc has a vanishing Lyapunov exponent along the
direction tangent to the arc itself, since neighboring points along the arc
are periodic orbits of the same order. On the other hand, in any
neighborhood of such a point there will be initial conditions at radii $%
r_{-} $ just above or below the arc that will not satisfy (\ref{arcs}) for
any values of $\ell$ and $k$. Such points will give rise to trajectories in
which the particle eventually does pass outside the interaction region, to
end up at the mercy of the alternating shears $T_{+}$ and $T_{-}$, which
give rise to the chaotic portions of the phase diagram. Thus, in general, we
expect arcs of periodic trajectories of this type to be largely immersed in
the chaotic parts of the phase diagram.

From the discussion above, which focuses on motions confined to the
interaction region, one may wonder whether there are counterparts to these
motions which take place with the particle confined entirely outside the
interaction zone. In fact, it is easy to convince oneself that such motions
are relatively few and far between. Indeed, for any point $X_{0}$ in $S_{%
\text{out}}$, the circle on which it moves necessarily intersects $D_{-}.$
Moreover, the angle $\delta_{+}\left( r_{+}\right) $ subtended by the two
intersection points of the edge of the disk $D_{-}$ and the circle of radius 
$r_{+}$ on which the phase point moves is now greater than $\pi$. Hence,
only period one orbits are able to avoid entering the interaction region.
Such fixed points will occur whenever the time the particle takes to
traverse the non-interaction region is an integer multiple of the oscillator
period.

\section{Other Periodic Orbits: Void II}

\label{voidTwo}

Having thus classified all orbits in which the particle never leaves the
interaction region, we now turn to the more complicated situation in which
the particle explores the entire configuration space available to it. Of
course, it is impossible to classify all such orbits, since they are
precisely the ones that are responsible for the variety of intricate
structures that appear, as well as for the chaos. Nonetheless, some of the
more prominent features of our figures can be explained quantitatively.
Indeed, we show in the analysis below that the fixed point at the centers of
the elliptic islands that we have referred to as Void II arise from a
relatively simple type of orbit in which the particle traverses each section
of the ring \emph{exactly once per period}. We also show that an infinite
number of such fixed points exists, and that most of them are hyperbolic,
and thus unstable.

\begin{figure}[t]
\centering \includegraphics[height=2.5in]{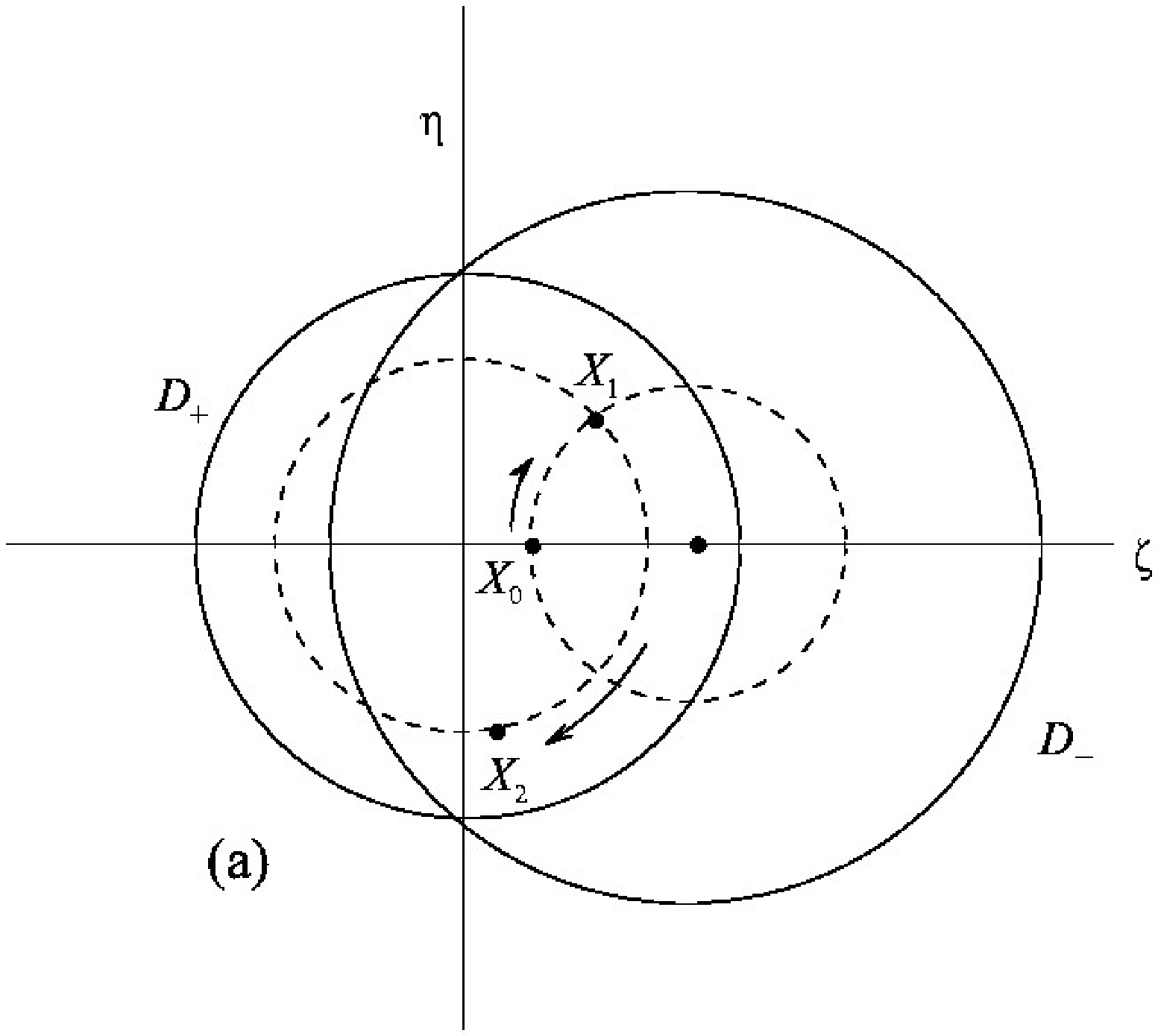} 
\hspace{0.50in}\centering %
\includegraphics[height=2.5in]{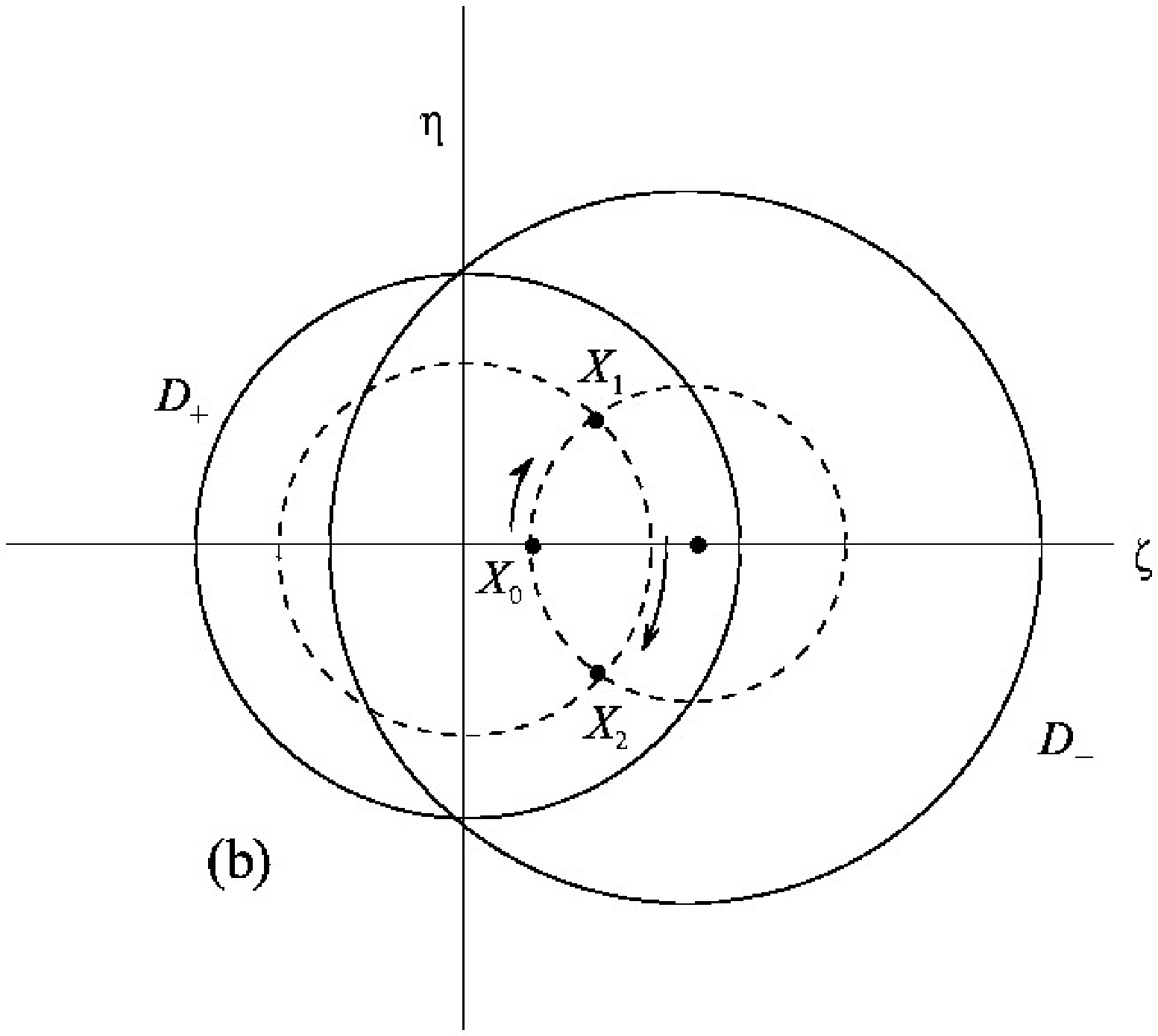} 
\caption{{Motion on the disks $D_{-}$ and $D_{+}$, of lozenge orbits
associated with fixed points of the dynamics.} }
\label{VoidII}
\end{figure}

Conceptually, such an orbit can be viewed within the geometric picture
developed above, as follows. Consider a phase point $X_{0}$ on the $\zeta $%
-axis within $D_{-}$ at an instant when the particle is at the center of the
interaction region and is moving to the right (See Fig.\ \ref{VoidII}(a)).
When the particle reaches $q=1,$ the oscillator phase point has rotated
through an angle $\varphi_{-}\left( r_{-}\right) /2$ about the center of $%
D_{-}$ to the point $X_{1},$ as shown. The particle now leaves the
interaction region and travels around the ring to $q=-1,$ while the
oscillator phase point rotates through an angle $\varphi_{+}\left(
r_{+}\right) $ to a point $X_{2} $. For the particle to now re-enter the
interaction region, the point $X_{2}$ must lie in the intersection region $S$%
, as shown in Fig.\ \ref{VoidII}(a). If, in addition, the system is to
return to its original state $X_{0}$ when the particle arrives again at $q=0,
$ it is obviously necessary for $X_{2}$ to fall on the circle of radius $%
r_{-}$ on which it started, as in Fig.\ \ref{VoidII}(b). Thus, all periodic
orbits that traverse each section of the ring exactly once per period have
vertices $X_{1}$ and $X_{2}$ that lie on a ``lozenge'' structure of this
type. Using relatively simple arguments, given below, we can locate the tops
of all such lozenge orbits and demonstrate in the process that they are
infinite in number.

\begin{figure}[t]
\centering\includegraphics[height=2.5in]{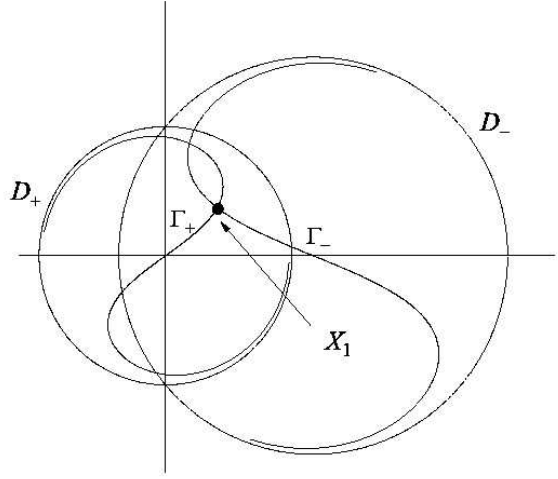} 
\caption{{Intersection of the curves $\Gamma_{-}$ and $\Gamma_{+}$},
described in the text, locating the tip $X_{1}$ of a lozenge orbit. Each
curve wraps an infinite number of times around the edge of the disk in which
it originates.}
\label{gamma}
\end{figure}

To proceed, we recall that any point $X_{1}$ which forms the top or bottom
of a lozenge orbit can be be obtained by rotating some point on the segment $%
\left( d-1,d+1\right) $ through an angle $\varphi_{-}\left( r_{-}\right) /2.$
The locus $\Gamma_{-}$ of all such points is depicted in Fig.\ \ref{gamma}
as a solid curve passing through the center of $D_{-}$. But any such point $%
X_{1}$ can also be obtained by rotating some point on the segment $\left( -%
\sqrt{1-d^{2}} ,\sqrt{1-d^{2}} \right) $ \emph{backwards in time} through an
angle $-\varphi_{+}\left( r_{+}\right) /2.$ The locus $\Gamma_{+}$ of this
set of points is depicted as a solid line passing through the center of $%
D_{+}$ in the same figure. It should now be clear that the intersection
points of these two curves locate all possible values $X_{1}$ associated
with periodic orbits of this type. For each $X_{1}$ so located, the
corresponding fixed point $X_{0}$ is obtained by rotating $X_{1}$ through $%
-\varphi_{-}\left( r_{-}\right) /2$. Since these two curves intersect an
infinite number of times, the number of such fixed points is, itself,
infinite. Indeed, the points $X_{1}^{\left( \infty\right) }$ and $%
X_{2}^{\left( \infty\right) }$ at the intersections of the edges of $D_{-}$
and $D_{+}$ are accumulation points of lozenge tips. In addition, any point
where either of the two curves plotted in Fig.\ \ref{gamma} crosses the edge
of the disk in which it did not originate will also be an accumulation point
of lozenge tips. As a result, it is clear that the points $X_{0\pm}^{\left(
\infty\right) }$ with coordinates $\left( 1\pm d,0\right) $ are accumulation
points for fixed points of the dynamics. As we will show shortly, and as one
might expect from the discussion of section \ref{chaos}, fixed points that
occur too near the edge of either disk (where the shear strengths diverge)
will be unstable.

On the other hand, the fixed points at the centers of the elliptic islands
visible in Figs.\ \ref{Eseries}(c)-(e), Figs.\ \ref{AlphaSeries}(a) and (c),
Figs.\ \ref{EseriesLow}(c), and Fig.\ \ref{Lseries}(b) and (c) are clearly
stable, and can therefore not be associated with an intersection of $%
\Gamma_{+}$ and $\Gamma_{-}$ that is too close to the edge of either disk.
Based on this argument, the most reasonable candidate for such a fixed point
is associated with the first intersection of $\Gamma_{+}$ and $\Gamma_{-}$
encountered when moving outward along $\Gamma_{+},$ starting from the
origin. This intersection is readily computed numerically. For values of $E,
L$, and $\alpha$ appropriate to our figures in which a Void II occurs, the
results of such a calculation are tabulated in Table \ref{FixedPoints}, and
are indicated as horizontal dashed lines in the associated figures. To
numerical accuracy the tabulated values agree with the actual locations of
the centers of the type II Voids in those figures, supporting the basic
picture developed above.

Let us now show how to predict in advance whether any particular
intersection of $\Gamma_{+}$ and $\Gamma_{-}$ gives rise to a stable or
unstable fixed point. To this end, it is clearly sufficient to consider the
stability of the evolution in the neighborhood of the upper lozenge tip. Let 
$\vec{x}_{1}$ and $\vec{x}_{2}$ be the vectors, respectively, from the
origin to the upper tip $X_{1}$, and to the lower tip $X_{2}$, of some
lozenge, and let $\vec{x}_{i}^{\prime}=\vec{x}_{i}-\vec{d}$ be the
corresponding vectors locating those points from the center of $D_{-}.$ By
definition, $\vec{x}_{2}=T_{+}\vec
{x}_{1}$, and $\vec{x}_{1}=T_{-}\vec{x}%
_{2},$ where $T_{+}$ and $T_{-}$ are the mappings associated with the
evolution of the system during one traversal of the corresponding region by
the particle. Note that $T_{-}T_{+}\vec{x}_{1}=\vec{x}_{1}$. It will
therefore be sufficient to compute the Jacobian matrix of $T_{-}T_{+}$ at $%
X_{1}$. Because the points $X_{1}$ and $X_{2}$ are obtained from one another
by moving along arcs of constant radius from the centers of $D_{\pm},$ it
follows that $r_{\pm}\left( \vec{x}_{1}\right) =r_{\pm}\left( \vec{x}%
_{2}\right) ,$ $\varphi_{+}\left( \vec{x}_{1}\right) =\varphi_{+}\left( \vec{%
x}_{2}\right) $ $=\phi,$ and $\varphi_{-}\left( \vec{x}_{1}\right)
=\varphi_{-}\left( \vec{x}_{2}\right) =\psi.$ Specifically, this means that%
\begin{equation*}
\vec{x}_{1}=\vec{d}+R\left( \psi\right) \left( \vec{x}_{2}-\vec{d}\right)
=T_{-}\;\vec{x}_{2}\qquad\vec{x}_{2}=R\left( \phi\right) \;\vec{x}_{1}=T_{+}%
\vec{x}_{1} 
\end{equation*}
where the rotation matrix 
\begin{equation*}
R\left( \theta\right) =\left( 
\begin{array}{cc}
\cos\theta & \sin\theta \\ 
-\sin\theta & \cos\theta%
\end{array}
\right) 
\end{equation*}
induces a rotation in the clockwise sense through an angle $\theta$. We
compute first the Jacobian matrix $J_{+}$ of $T_{+}$ at $X_{1}$%
\begin{align*}
J_{+} & =\left( 
\begin{array}{cc}
\cos\phi & \sin\phi \\ 
-\sin\phi & \cos\phi%
\end{array}
\right) +\frac{\phi^{\prime}}{r_{+}}\left( 
\begin{array}{cc}
-\sin\phi & \cos\phi \\ 
-\cos\phi & -\sin\phi%
\end{array}
\right) \left( 
\begin{array}{cc}
x_{1}x_{1} & x_{1}y_{1} \\ 
x_{1}y_{1} & y_{1}y_{1}%
\end{array}
\right) \\
& =\left( 
\begin{array}{cc}
\cos\phi & \sin\phi \\ 
-\sin\phi & \cos\phi%
\end{array}
\right) -\frac{\phi^{\prime}}{r_{+}}\left( 
\begin{array}{cc}
\cos\left( \phi-\pi/2\right) & \sin\left( \phi-\pi/2\right) \\ 
-\sin\left( \phi-\pi/2\right) & \cos\left( \phi-\pi/2\right)%
\end{array}
\right) \left( 
\begin{array}{cc}
x_{1}x_{1} & x_{1}y_{1} \\ 
x_{1}y_{1} & y_{1}y_{1}%
\end{array}
\right) \\
& =R\left( \phi\right) - r_{+}\phi^{\prime}R\left( \phi-\pi/2\right) \;\hat{x%
}_{1}\hat{x}_{1}
\end{align*}
where $\hat{x}\hat{x}$ denotes the second rank tensor product of the unit
vector $\hat{x}$ with itself. Similarly, we consider the Jacobian $J_{-}$ of 
$T_{-}$ at $X_{2}$ 
\begin{align*}
J_{-} & =\left( 
\begin{array}{cc}
\cos\psi & \sin\psi \\ 
-\sin\psi & \cos\psi%
\end{array}
\right) +\frac{\psi^{\prime}}{r_{-}\left( \vec{x}_{2}\right) }\left( 
\begin{array}{cc}
-\sin\psi & \cos\psi \\ 
-\cos\psi & -\sin\psi%
\end{array}
\right) \left( 
\begin{array}{cc}
x_{2}^{\prime}x_{2}^{\prime} & x_{2}^{\prime}y_{2} \\ 
x_{2}^{\prime}y_{2} & y_{2}^{\prime}y_{2}^{\prime}%
\end{array}
\right) \\
& =\left( 
\begin{array}{cc}
\cos\psi & \sin\psi \\ 
-\sin\psi & \cos\psi%
\end{array}
\right) -\frac{\psi^{\prime}}{r_{-}\left( \vec{x}_{2}\right) }\left( 
\begin{array}{cc}
\cos\left( \psi-\pi/2\right) & \sin\left( \psi-\pi/2\right) \\ 
-\sin\left( \psi-\pi/2\right) & \cos\left( \psi-\pi/2\right)%
\end{array}
\right) \left( 
\begin{array}{cc}
x_{2}^{\prime}x_{2}^{\prime} & x_{2}^{\prime}y_{2} \\ 
x_{2}^{\prime}y_{2} & y_{2}^{\prime}y_{2}^{\prime}%
\end{array}
\right) \\
& =R\left( \psi\right) -r_{-}\psi^{\prime}R\left( \psi-\pi/2\right) \hat{x}%
_{2}^{\prime}\hat{x}_{2}^{\prime}.
\end{align*}
The trace of the product%
\begin{align*}
J_{-}J_{+} & =(R\left( \psi\right) -r_{-}\psi^{\prime}R\left( \psi
-\pi/2\right) \hat{x}_{2}^{\prime}\hat{x}_{2}^{\prime})(R\left( \phi\right)
-r_{+}\phi^{\prime}R\left( \phi-\pi/2\right) \;\hat{x}_{1}\hat{x}_{1}) \\
& =R\left( \phi+\psi\right) -r_{-}\psi^{\prime}R\left( \psi-\pi/2\right) 
\hat{x}_{2}^{\prime}\hat{x}_{2}^{\prime}R\left( \phi\right)
-r_{+}\phi^{\prime}R\left( \psi+\phi-\pi/2\right) \;\hat{x}_{1}\hat{x}_{1} \\
& \ \qquad\qquad\qquad\qquad\qquad+ r_{-}r_{+}\psi^{\prime}\phi^{\prime
}R\left( \psi-\pi/2\right) \hat{x}_{2}^{\prime}\hat{x}_{2}^{\prime}R\left(
\phi-\pi/2\right) \;\hat{x}\hat{x}
\end{align*}
\newline
of these two matrices determines the stability of the evolution in the
neighborhood of the lozenge tip $X_{1}$ associated with the orbit. The trace
of the first term is obvious. To evaluate the second, use $\text{Tr}\left[ 
\vec{x}\vec{y}\right] =\vec{x}\cdot\vec{y},$ and the invariance of the trace
of a product of operators under their cyclic permutation: 
\begin{align*}
\text{Tr}R\left( \psi-\pi/2\right) \hat{x}_{2}^{\prime}\hat{x}_{2}^{\prime
}R\left( \phi\right) & =\text{Tr}R\left( \phi\right) R\left( \psi
-\pi/2\right) \hat{x}_{2}^{\prime}\hat{x}_{2}^{\prime}=\text{Tr}\left(
R\left( \phi+\psi-\pi/2\right) \hat{x}_{2}^{\prime}\right) \cdot\hat{x}%
_{2}^{\prime} \\
& =\cos\left( \phi+\psi-\pi/2\right) .
\end{align*}
Similarly, $\text{Tr}\left[ R\left( \psi+\phi-\pi/2\right) \;\hat{x}_{1}\hat{%
x}_{1}\right] =\cos\left( \phi+\psi-\pi/2\right) .$ The trace of the last
term similarly reduces to the product%
\begin{align*}
\text{Tr}\left[ R\left( \psi-\pi/2\right) \hat{x}_{2}^{\prime}\hat{x}%
_{2}^{\prime}\;R\left( \phi-\pi/2\right) \hat{x}_{1}\hat{x}_{1}\right] &
=\left( \hat{x}_{2}^{\prime}\cdot R\left( \phi-\pi/2\right) \hat{x}%
_{1}\right) \text{Tr}\left[ R\left( \psi-\pi/2\right) \hat{x}_{2}^{\prime }%
\hat{x}_{1}\right] \\
& =\left( \hat{x}_{2}^{\prime}\cdot\left[ R\left( \phi-\pi/2\right) \hat{x}%
_{1}\right] \right) \left( \hat{x}_{1}\cdot\left[ R\left( \psi -\pi/2\right) 
\hat{x}_{2}^{\prime}\right] \right) .
\end{align*}
To evaluate this we need to consider more carefully the geometry of the
situation. Let $\theta_{-}=$ $\psi/2$ and $\theta_{+}=\phi/2\ $be the
interior angles adjacent to the base of the triangle connecting the centers
of $D_{-}$ and $D_{+}$ to the point $X_{1}.$ Note that the sum $%
\theta_{+}+\theta_{-}\leq\pi$. The angle $\chi=\pi-\theta_{+}-\theta_{-}$
opposite this base is the angle between the vectors $\vec{x}_{1}$ and $\vec{x%
}_{1}-\vec{d}=\vec{x}_{1}^{\prime}.$ So%
\begin{equation*}
\hat{x}_{1}=R\left( \pi-\theta_{+}-\theta_{-}\right) \hat{x}_{1}^{\prime} . 
\end{equation*}
But a vector along $\hat{x}_{2}^{\prime}$ is rotated into a vector along $%
\hat{x}_{1}^{\prime}$ by a rotation through $\phi=2\theta_{-},$ i.e., $\hat{x%
}_{1}^{\prime}=R\left( 2\theta_{-}\right) \hat{x}_{2}^{\prime},$ so%
\begin{equation*}
\hat{x}_{1}=R\left( \pi-\theta_{+}-\theta_{-}\right) R\left( 2\theta
_{-}\right) \hat{x}_{2}^{\prime}=R\left( \pi-\theta_{+}+\theta_{-}\right) 
\hat{x}_{2}^{\prime}. 
\end{equation*}
This leads to%
\begin{equation*}
\hat{x}_{1}\cdot\left[ R\left( \psi-\pi/2\right) \hat{x}_{2}^{\prime }\right]
=\hat{x}_{1}\cdot\left[ R\left( \frac{\psi}{2}+\frac{\phi}{2}-3\pi/2\right) 
\hat{x}_{1}\right] =\cos\left( \frac{\psi}{2}+\frac{\phi}{2}-3\pi/2\right)
=-\sin\left( \frac{\psi+\phi}{2}\right) 
\end{equation*}
and%
\begin{align*}
\hat{x}_{2}^{\prime}\cdot\left[ R\left( \phi-\pi/2\right) \hat{x}_{1}\right]
& =\hat{x}_{2}^{\prime}\cdot\left[ R\left( \phi-\pi/2\right) R\left(
\pi-\theta_{+}+\theta_{-}\right) \hat{x}_{2}^{\prime}\right] =\hat{x}%
_{2}^{\prime}\cdot\left[ R\left( \frac{\psi}{2}+\frac{\phi}{2}+\pi/2\right) 
\hat{x}_{2}^{\prime}\right] \\
& =\cos\left( \frac{\psi}{2}+\frac{\phi}{2}+\pi/2\right) =-\sin\left( \frac{%
\psi+\phi}{2}\right) .
\end{align*}
Thus, for the lozenge evolution,%
\begin{equation}
\text{Tr}\left[ J_{-}J_{+}\right] =2\cos\left( \psi+\phi\right) -\left(
r_{-}\psi^{\prime}+r_{+}\phi^{\prime}\right) \sin\left( \phi+\psi\right)
+r_{-}r_{+}\psi^{\prime}\phi^{\prime}\sin^{2}\left( \frac{\psi+\phi}{2}%
\right)
\end{equation}
or%
\begin{equation}
\text{Tr}\left[ J_{-}J_{+}\right] =2+\left(
r_{-}r_{+}\psi^{\prime}\phi^{\prime}-4\right) \sin^{2}\left( \frac{\psi+\phi%
}{2}\right) -2\left( r_{-}\psi^{\prime}+r_{+}\phi^{\prime}\right) \sin\left( 
\frac{\psi+\phi}{2}\right) \cos\left( \frac{\psi+\phi}{2}\right) .
\end{equation}
From this form it is first of all possible to prove, confirming our previous
arguments, that fixed points associated with lozenge tips close to the edge
of either disk (where $\psi^{\prime}$ and $\phi^{\prime}$ diverge) are
unstable. This is easily shown for $d<1/\sqrt{2}$. In that case, a little
trigonometry shows that the angle $\chi$ associated with any lozenge tip
near the boundary of $S$ is acute, so that $\pi\geq\left( \psi+\phi\right)
/2>\pi/2.$ Consequently, $\text{Tr}\left[ J_{-}J_{+}\right] \geq2$, with
equality only if $\sin\left( \psi+\phi\right) /2=0$, so that the fixed point
is hyperbolic or, at worst, parabolic. For this last case to occur the
angles $\psi/2$ and $\phi/2$ must add up to $\pi,$ which can only happen if $%
\Gamma_{+}$ and $\Gamma_{-}$ intersect on the horizontal axis. In that case, 
$\phi/2=\pi$ and $\psi/2=0,$ or vice versa. It is not hard to convince
oneself that, for any fixed value of $d,$ there exist arbitrarily large
values of $a_{-}$ and $a_{+}$ where such fixed points will appear. All other
fixed point of this type with $d<1/\sqrt{2}$ are, however, hyperbolic.

\begin{table}[t]
\begin{equation*}
\begin{tabular}{lcc}
\toprule Figure & \ \ $\zeta_{0} -d$\ \  & \ \ Tr$[J_{-}J_{+}]$ \ \  \\ 
\colrule{\ref{Eseries}}(c) & $-0.546$ & $-1.995$ \\ 
{\ref{Eseries}}(d) & $-0.389$ & $-0.849$ \\ 
{\ref{Eseries}}(e) & $-0.306$ & $-0.078$ \\ 
{\ref{AlphaSeries}}(a) & $-0.346$ & $-1.976$ \\ 
{\ref{AlphaSeries}}(b) & $-0.053$ & $-1.995$ \\ 
{\ref{AlphaSeries}}(b) & $+0.878$ & $-0.948$ \\ 
{\ref{AlphaSeries}}(b) & $-0.864$ & $+5.812$ \\ 
{\ref{AlphaSeries}}(c) & $-0.017$ & $-1.995$ \\ 
{\ref{EseriesLow}}(b) & $-0.575$ & $-2.026$ \\ 
{\ref{EseriesLow}}(c) & $-0.392$ & $-0.363$ \\ 
{\ref{Lseries}(b)} & $-0.514$ & $-1.704$ \\ 
{\ref{Lseries}(c)} & $-0.224$ & $+0.384$ \\ 
\botrule &  & 
\end{tabular}
\end{equation*}
\caption{Location and stability of some lozenge type fixed points $X_{0}=(%
\protect\zeta_{0},\protect\eta_{0})=(\protect\zeta_{0},0)$ appearing in the
figures indicated.}
\label{FixedPoints}
\end{table}

Using the algorithm described above we have calculated numerically the
location and stability, as indicated by the value of $\text{Tr}\left[
J_{-}J_{+}\right] $, for the first intersection of $\Gamma_{+}$ and $%
\Gamma_{-}$ for the system parameters of a certain number of our figures.
The results appear in Table \ref{FixedPoints}. Comparison with the relevant
figures shows that whenever $|\text{Tr}\left[ J_{-}J_{+}\right] |<2$ the
associated fixed point is indeed the center of a Void II. For Fig. \ref%
{EseriesLow}(b), this fixed point, computed to lie at $\zeta- d = -0.575$,
has a trace with magnitude just greater than 2. It appears as the hyperbolic
structure located below Void I in that figure. In Fig.\ \ref{AlphaSeries}%
(b), we have, in addition, computed the second and third intersection of $%
\Gamma_{+}$ and $\Gamma_{-}$. One leads to a fixed point at $\zeta- d = 0.878
$, which is stable, and appears at the center of a crescent shaped strucure
at the top of that figure. The other occurs at $\zeta- d = -0.864$, but is
unstable. It does not, therefore, give rise to an elliptic island, but lies
right at the edge of Void II.

Finally, although in general the location of the fixed point at the center
of Void II must be determined numerically, its behavior for large energy can
be obtained from the limiting behavior of the dynamics as $d,$ $a_{+},$ and $%
a_{-}$ go to zero. In this limit, denoting the location of the fixed point
as $X_{0}=\left( \eta d,0\right) ,$ we note that the curves $\Gamma_{\pm}$
are well represented by straight lines in the neighborhood of the origin.
The corresponding triangle bounded by the segment $\left[ 0,d\right] $ and $%
\Gamma_{\pm}$ then has a height $h=\eta d\tan a_{+}\sim\eta da_{+}=\eta
Ld^{2}/\alpha$ as measured from the origin, and $h=\left( 1-\eta\right)
d\tan a_{-}\sim\left( 1-\eta\right) da_{-}=\left( 1-\eta\right) d^{2}/\alpha$
as measured from $\left( d,0\right) ,$ so that%
\begin{equation}
\eta d=\frac{2d}{2+L}   \label{HighEFixedPoint}
\end{equation}
locates the corresponding fixed point on the segment $\left[ 0,d\right] .$
This has a simple physical interpretation. At high energies, when the
particle is moving very fast, the fractional change in its speed as it
enters and exits the interaction region is small. In such a motion, the
reduced oscillator coordinate oscillates about equilibrium position $d$
during that fraction $2/\left( 2+L\right) $ of the time the particle is in
the interaction region, and oscillates about the origin during that fraction 
$L/\left( 2+L\right) $ of the time that the particle is outside the
interaction region. A time average of these two values of the equilibrium
position results in the location (\ref{HighEFixedPoint}) of the fixed point
in this high energy limit.

Having thus made our point that the most prominent features that appear in
our phase plots can be explained in terms of the simplest periodic orbits of
the system, we point out that many other periodic orbits occur in which the
particle traverses the two sections of the ring more than once per period.
Such orbits will then give rise to the more complicated structures
appearing, e.g. in Fig.\ \ref{pigface}. We conjecture that a more complete
analysis of those orbits along lines similar to those developed in this
paper, would allow a complete explanation of some of the more
anthropomorphic structures appearing in that figure.

\section{Discussion and Summary}

\label{discussion}

The model we have introduced and studied in this paper
clearly has certain similarities to a number of previously studied dynamical
systems, such as the kicked rotor, the Fermi accelerator, billiards, and the
spring-pendulum. Indeed, the study of the dynamics reduces in all these
cases to the study of a return map on a suitably chosen Poincar\'e section.
The model presented here nevertheless differs in important ways from each of
these previously studied systems.

Indeed, the kicked rotor and the Fermi accelerator describe externally
perturbed nonconservative systems with one degree of freedom, while the
current model is closed, energy-conserving, and has two degrees of freedom.
In our model, although the particle gets ``kicked'' each time it reaches 
$q=\pm 1$, the kicks are neither periodic in time, as in the rotor, nor are
they imposed by an external agent, as is the case for both the rotor and the
Fermi accelerator.

Billiards, of course, are closed conservative systems with two degrees of
freedom as well, but trajectories in billiards have the unusual property of
being independent of particle energy, so that changing the energy does not
change the statistical features of the dynamics. This is in sharp contrast
to the behavior of the present model, in which many of the interesting
features that arise, do so as a result of changes in the energy of the
system under conditions in which the underlying potential is kept fixed.
Moreover, the present system can be interpreted in terms of the Hamiltonian
interaction between two otherwise separate mechanical systems. It should,
therefore, be of more use in understanding many problems for which that is
an essential feature.

Another closed Hamiltonian system of that type is the spring-pendulum \cite%
{Broer1,Broer2}. Like the current model, the spring-pendulum is a
Hamiltonian system with two degrees of freedom, each of which has a simple
description when treated on its own. Unlike the spring-pendulum, however,
the present model has the advantage that the coupling between the two
sub-systems can be smoothly turned off in a way that allows the unperturbed
dynamics of each to be recovered, and thus may be more useful for
understanding fundamental properties of interacting independent systems of
this type. We note that a bifurcation analysis for the spring-pendulum was
given in \cite{Broer2}. 
A similar analysis applied to our model would yield predictions on the
motion and the shape of the elliptic islands that appear in our model as the
parameters are changed, and could be an interesting direction for future
study. Our emphasis here has instead been on some unusual features of our
model, that we now briefly recall.

We have argued that for suitable system parameters the system exhibits
either a fully chaotic phase space, or a mixed phase space in which only two
regions occur. In one the motion is chaotic, and in the other it is
completely integrable with no secondary KAM structures (See Fig.\ \ref%
{Eseries}(b) and (c)). While we have explained the presence of chaos in the
model in terms of alternating shears similar to the kind that arise in
linked twist maps, our conjecture regarding the absence of secondary
structures of finite measure in the chaotic sea for small energies is based
largely on numerical calculations over very small regions of phase space
that have heretofore failed to detect any structure in the chaotic region.
We have not, however, given a rigorous proof that such structures do not
emerge at small enough length scales in phase space.

We note also that the sharp boundary that exists in this model between the
chaotic and the completely integrable parts of phase space should make it an
interesting system on which to test current conjectures of quantum chaos
theory, in particular those pertaining to systems with a mixed phase space,
and to the localization properties of eigenfunctions on chaotic and
completely integrable parts of phase space. While the clear-cut boundary of
the present model should facilitate such an analysis, \cite{MP,marklof} it
does have the complication of an underlying potential that has finite
discontinuities in configuration space, and the matching problems that
occur, in contrast, e.g., to billiard systems, for which a number of
efficient numerical techniques have been developed for solving the
corresponding Schr\"{o}dinger equation \cite{baecker}.

Our own interest in the present system arose originally from a fundamental
interest in Hamiltonian models of transport and dissipation in deformable
media. In many such systems, the medium through which a particular transport
species moves can be modeled as an appropriate collection of harmonic
oscillators. Among the many questions that arise in such extended systems
is, e.g., to what extent the presence of microscopic chaos, or its absence,
in the interaction of a particle with a single oscillator, manifests itself
in the transport properties that emerge at long times after repeated
interactions with many independent or mutually-coupled oscillators. We view
the present analysis as a step towards answering this and other questions
related to systems of this kind.

\begin{acknowledgments}
{This work was supported in part by the NSF under grants DMR-0097210 and
INT-0336343. SDB acknowledges the hospitality of and the MSRI-Berkeley, and
PEP that of the Universit\'{e} des Sciences et Technologies de Lille, and
the Consortium of the Americas for Interdisciplinary Science, University of
New Mexico, where part of this work was performed.}
\end{acknowledgments}

\end{document}